\RequirePackage{lineno}
\documentclass[prb,twocolumn,showpacs,amsbsy,linenumbers,floatfix,superscriptaddress]{revtex4}
\usepackage{epsfig,color}
\usepackage[cp1250]{inputenc}
\usepackage{amsmath}
\begin{document}

\title{Tuning of the spin-orbit interaction in a quantum dot\\ by an in-plane magnetic field}

\author{M.P. Nowak}
\affiliation{Faculty of Physics and Applied Computer Science, AGH University of Science and Technology, \\
al. Mickiewicza 30, 30-059 Krak\'ow, Poland}
\affiliation{Departement Fysica, Universiteit Antwerpen, Groenenborgerlaan 171,
  B-2020 Antwerpen, Belgium}
\author{B. Szafran}
\affiliation{Faculty of Physics and Applied Computer Science, AGH University of Science and Technology, \\
al. Mickiewicza 30, 30-059 Krak\'ow, Poland}
\author{F.M. Peeters}
\affiliation{Departement Fysica, Universiteit Antwerpen, Groenenborgerlaan 171,
  B-2020 Antwerpen, Belgium}
\author{B. Partoens}
\affiliation{Departement Fysica, Universiteit Antwerpen, Groenenborgerlaan 171,
  B-2020 Antwerpen, Belgium}
\author{W. Pasek}
\affiliation{Faculty of Physics and Applied Computer Science, AGH University of Science and Technology, \\
al. Mickiewicza 30, 30-059 Krak\'ow, Poland}

\date{\today}

\begin{abstract}

Using an exact diagonalization approach we show that one- and two-electron InAs quantum dots exhibit avoided crossing in the energy spectra that are induced by the spin-orbit coupling in the presence of an in-plane external magnetic field. The width of the avoided crossings depends strongly on the orientation of the magnetic field which reveals the intrinsic anisotropy of the spin-orbit coupling interactions. We find that for specific orientations of the magnetic field avoided crossings vanish. Value of this orientation can be used to extract the ratio of the strength of Rashba and Dresselhaus interactions. The spin-orbit anisotropy effects for various geometries and orientations of the confinement potential are discussed. Our analysis explains the physics behind the recent measurements performed on a gated self-assembled quantum dot [S. Takahashi {\it et al.} Phys. Rev. Lett. {\bf 104}, 246801 (2010)].
\end{abstract}

\maketitle

\section{Introduction}
Over the past decade there has been a growing interest in the study of the spin-orbit (SO) interaction in semiconductor low-dimensional systems motivated by the possibility of coherent spin manipulation.\cite{nowack,qubdrut,decay,wal,psh,nitta,bso,bedn,sfet,nonsfet,anisotropic,ldqg,aour} The Hamiltonians describing the SO coupling resulting from the inversion asymmetry of the material (Dresselhaus\cite{dress} coupling) or the specific structure of the device (Rashba\cite{rashba} interaction) are not invariant with respect to the rotation of the spin or the momentum operators separately, and consequently spin-orbit-coupled systems posses intrinsic anisotropic properties. This anisotropy has been thoroughly studied for delocalized systems.\cite{wal,psh} In particular in transport experiments the dependence of the conductance of a narrow quantum wire on the direction of the external magnetic field can be used to determine the reciprocal strengths of the Rashba and Dresselhaus couplings.\cite{nitta} The anisotropy of the spin-orbit interaction is translated into anisotropic effective magnetic field\cite{bso} for a moving electron modifying the electron spin state. This effective magnetic field can be used to perform rotations of spin and thus to construct quantum gates \cite{bedn} or a spin-field effect transistor.\cite{sfet,nonsfet} Moreover, the spin-orbit coupling is responsible for anisotropic corrections \cite{anisotropic} to the spin swap in a two-qubit quantum gate,\cite{ldqg} because it results into the precession of spin-packets tunneling between the two quantum dots.\cite{aour}

For electrons localized in a quantum dot the SO coupling results in avoided crossings (AC) in the energy spectra\cite{acac} and spin relaxation\cite{srelax} mediated by phonons with a relaxation rate dependent on the orientation of the external magnetic field.\cite{asrelax} The energetic effects of the SO interaction are usually weak. Only recently SO-induced AC were experimentally measured on quantum-dots that were situated in gated nano-wires \cite{whisk1,whisk2} and in gated self-assembled quantum dots.\cite{tarucha} The latter experiment studied changes of the width of AC for different orientations of the magnetic field which extended the previous studies that were focused on a comparison of the spin-splittings for vertical and in-plane alignment of magnetic field \cite{prevexp,oblicz1} in circularly symmetric confinement potentials.

In the present work we explain the physics underlying the observations of Ref. \onlinecite{tarucha}. To the best of our knowledge the present paper explains for the first time the oscillatory dependence of the width of AC on the direction of the in-plane magnetic field. The latter turns out to be the consequence of the influence of the individual SO couplings and the anisotropy of the confinement potential. This conclusion is supported by an exact three-dimensional calculation of the energy spectra of one- and two-electron spin-orbit-coupled quantum dots.

We show that for quantum dots with confinement potential elongated in $[100]$ direction for pure Rashba (pure Dresselhaus) coupling the AC disappears when the magnetic field is aligned along the short (long) axis of the dot. We show how this can be understood from the form of the SO Hamiltonians and the approximate parity of the one-electron wave functions. The dependence of the AC width on the direction of the magnetic field turns to be a $|\sin\phi|$ shaped function and when both couplings are present this function is shifted by an amount which depends on the relative strength of both interactions. This shift is affected by the orientation of the dot within the [001] plane due to the SO bulk-induced-anisotropy (Dresselhaus term).  For completeness we also study the influence of the dot shape. We show that for a square-based quantum dot the anisotropic dependence of the AC width is only observed when both couplings are present.\cite{oblicz1} Moreover we show that for increased height of the dot the orbital effect of the magnetic field modifies the energy spectrum but the shape of the dependence of the anticrossing width on the direction of the in-plane magnetic field remains unaltered.

The present work is organized as follows: we start with an outline of our theoretical approach in section II. In section III we present our numerical results starting from the single-electron case which provides us with physical insight in the reasons for the SO coupling anisotropy. We continue by studying different orientations and geometries of the dot and we end the section with the two-electron case that allows for a direct comparison with the recent experimental data of Ref. \onlinecite{tarucha}. We end with a concluding discussion in section IV and a summary in section V.

\section{Theory}

\subsection{Model}
\begin{figure}[ht!]
\epsfysize=60mm
                \epsfbox[20 168 561 681] {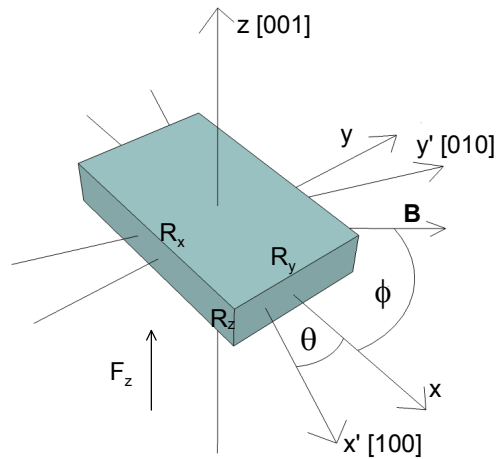}
                 \caption{Schematic of the quantum dot system with the used coordinate system fixed to the quantum dot. The crystallographic directions of the InAs host lattice are also indicated together with the direction of the in-plane magnetic field.}
 \label{soqd}
\end{figure}

Our aim is to calculate the energy spectra of the one- and two-electrons confined in a three-dimensional quantum dot in the presence of SO coupling and a magnetic field oriented within the quantum dot plane. The effect of the spin-orbit coupling on the energy are very small requiring a very high numerical precision when evaluating the energy spectrum. We assume that the quantum dot is cuboid in shape and that the confinement potential is separable, namely $V(\mathbf{r})=V_x(x)+V_y(y)+V_z(z)$. Moreover, we assume that the one-dimensional confinement potentials $V_x, V_y$ and $V_z$ can be described by an infinite quantum well model. This is a reasonable approximation for not to small quantum dots. Under these assumptions one can construct a sufficiently precise solver for the two-electron problem. We consider quantum dot with varied in-plane orientation with respect to the crystal host. The $z$-axis is taken along the $[001]$ crystal direction which is also the vertical axis of the dot. The orientation of the dot is described by a rotations of the $x$ and $y$ directions, which are the axes of the dot with respect to the $[100]$ and $[010]$ crystal directions. The outline of our quantum dot and the coordinate system used is depicted in Fig. \ref{soqd}.

\subsection{Method}

We employ the effective mass approximation with the single-electron Hamiltonian of the form

\begin{equation}
h=\left(\frac{\hbar^2\textbf{k}^2}{2m^*}+V(\textbf{r})\right)\textbf{1} + \frac{1}{2}g\mu_B \textbf{B}\cdot\mathbf{\sigma} + H_{BIA}+ H_{SIA},\label{h1e}
\end{equation}
where $\textbf{k}=-i\nabla+\frac{e\textbf{A}}{\hbar}$, $\textbf{1}$ is the identity matrix, $V(\textbf{r})$ defines the confining potential and $H_{BIA}$, $ H_{SIA}$ are the spin-orbit coupling Hamiltonians. The $x,y$ and $z$ directions are taken along the axes of the dot. But notice that the SO interaction Hamiltonians are defined in the coordinate system with axes parallel to the $[100]$, $[010]$ and $[001]$ which we denote with $x', y'$ and $z$. Both coordinate systems are transformed into each other by an in-plane rotation over an angle $\theta$.

We introduce the Rashba coupling with Hamiltonian,
\begin{equation}
H_{SIA}=\alpha \nabla' V \cdot (\sigma' \times \mathbf{k}'),\label{hasia}
\end{equation}
where $\alpha$ defines the coupling strength. For infinite quantum well confinement the term $\nabla' V$ within the dot equals the external electric field. We neglect the influence of the in-plane component of the electric field\cite{neglfy} and obtain the Rashba Hamiltonian in the form,

\begin{equation}
H_{SIA}=\alpha\left[ \frac{\partial V}{\partial z}\right] (\sigma_{x'} k_{y'} - \sigma_{y'} k_{x'}).\label{hasiafz}
\end{equation}
Thus the electric field is in the $z$ direction which is incorporated by taking a non-zero slope of the bottom of $V_z(z)$.

Inversion asymmetry of the crystal lattice results in a Dresselhaus SO coupling that is described by the Hamiltonian

\begin{equation}
\begin{split}
  H_{BIA} = \gamma &\left[ \sigma_{x'} k_{x'} (k^2_{z} - k^2_{y'}) + \sigma_{y'}k_{y'}(k^2_{x'} - k^2_{z})\right.\\
  & \left.+ \sigma_{z}k_{z}(k^2_{y'} - k^2_{x'})\right],\label{habia3d}
\end{split}
\end{equation}
where $\gamma$ is the coupling constant.

The coordinate system used for the SO coupling can be transformed into the coordinate system used for the quantum dot through the transformation

\begin{equation}
\begin{split}
x'=x\cos(\theta)-y\sin(\theta)\\
y'=x\sin(\theta)+y\cos(\theta)\\
\end{split}
\end{equation}
which applies both to the Pauli matrices $\sigma$ and the coordinates of the momentum operator.

We include an in-plane magnetic field of orientation $\textbf{B}=B\left(\cos\phi,\sin\phi,0\right)$ which is described by the gauge $\textbf{A}=B\left(z\cdot\sin\phi, 0,y\cdot\cos\phi\right)$. The magnetic field vector $\mathbf{B}$ for $\phi=0$ is oriented along the $x$ direction (see Fig. \ref{soqd}).

The one-electron Hamiltonian (\ref{h1e}) can be rewritten in the form $h=h_x+h_y+h_z+h_{ns}$, where

\begin{equation}
h_x=-\frac{\hbar^2}{2m^*}\frac{\partial^2}{\partial x^2}+V_x(x),\label{hs1}
\end{equation}

\begin{equation}
h_y=-\frac{\hbar^2}{2m^*}\frac{\partial^2}{\partial y^2} + V_y(y) + \frac{e^2B^2}{2m^*}y^2\cos^2\phi,\label{hs2}
\end{equation}

\begin{equation}
h_z=-\frac{\hbar^2}{2m^*}\frac{\partial^2}{\partial z^2} + V_z(z) + \frac{e^2B^2}{2m^*}z^2\sin^2\phi,\label{hs3}
\end{equation}
are spin independent parts separable in the $x,y$ and $z$ directions, and

\begin{equation}
\begin{split}
h_{ns}=-\frac{i\hbar eB}{m^*}\left(z\sin\phi\frac{\partial}{\partial x}+y\cos\phi\frac{\partial}{\partial z}\right) \\
+\frac{1}{2}g\mu_bB\left[ \sigma_x \cos\phi + \sigma_y\sin\phi \right] + H_{SIA} + H_{BIA},
\end{split}
\end{equation}
is the nonseparable part that contains the spin dependent terms,

The eigenenergies and the eigenvectors $\psi_x(x),\psi_y(y),\psi_z(z)$ of the Hamiltonians $h_x, h_y, h_z$ are calculated separately on one-dimensional meshes of $N_{1\mathrm{D}}=1000$ points. In a next step we diagonalize $h_{ns}$ in a basis of products of the eigenstates $\psi_x(x),\psi_y(y),\psi_z(z)$ resulting in three--dimensional spin-orbitals $\psi(\mathbf{r},\sigma)$. We typically take $N_x=N_y=20$, $N_z=10$ one-dimensional eigenstates (we assumed $R_z \ll R_x, R_y$), which including the degeneracy of the spin gives a basis consisting of $8000$ elements which results in an accuracy better than $5\mu$eV.

We solve the two-electron problem as described by the Hamiltonian

\begin{equation}
H=h_1 + h_2 + \frac{e^2}{4\pi\varepsilon\varepsilon_0|\mathbf{r}_1-\mathbf{r}_2|}\label{h2e},
\end{equation}
using the configuration-interaction approach. In our numerical calculation we take the dielectric constant for InAs $\varepsilon=14.6$. Hamiltonian (\ref{h2e}) is diagonalized in a basis constructed of antisymmetrized single-electron spin-orbitals $\psi(\mathbf{r},\sigma)$

 \begin{equation}
\Psi = \frac{1}{\sqrt{2}}\sum_{i=1}^n\sum_{j=i+1}^n\left[ \psi_i(\mathbf{1})\psi_j(\mathbf{2}) -\psi_i(\mathbf{2}) \psi_j(\mathbf{1})  \right],
\end{equation}
where $\mathbf{1},\mathbf{2}$ are the spatial $(\mathbf{r})$ and spin $(\sigma)$ coordinates of the corresponding electron. The electron-electron interaction matrix element requires the calculation of integrals of the form,

\begin{equation}
\begin{split}
&\frac{e^2}{4\pi\varepsilon_0}\langle\psi_i(\mathbf{r}_1)\psi_j(\mathbf{r}_2)|\frac{1}{\varepsilon |\mathbf{r}_1-\mathbf{r}_2|}|\psi_k(\mathbf{r}_1)\psi_l(\mathbf{r}_2)\rangle=\\
&e\int d^3\mathbf{r}_1 \psi_i^*(\mathbf{r}_1) \psi_k(\mathbf{r}_1) \int d^3\mathbf{r}_2 \frac{e}{4\pi\varepsilon_0}\frac{\psi_j^*(\mathbf{r}_2) \psi_l(\mathbf{r}_2)}{\varepsilon |\mathbf{r}_1-\mathbf{r}_2|}=\\
&e\int d^3\mathbf{r}_1 \psi_i^*(\mathbf{r}_1) \psi_k(\mathbf{r}_1) V_{jl}(\mathbf{r}_1).
\end{split}
\end{equation}
A direct calculation of these $6$ dimensional integrals requires an enormous numerical cost. Therefore, we use the method\cite{stopa} in which the innermost integral is attributed to an electric potential $V_{jl}(\mathbf{r}_1)$ originating from an electric charge distribution $\psi_j^*(\mathbf{r}_2) \psi_l(\mathbf{r}_2)$. We calculate the electric potential by solving the Poisson equation $\nabla^2V_{jl}(\mathbf{r}_1)=-e/(\varepsilon\varepsilon_0)\psi_j^*(\mathbf{r}_1) \psi_l(\mathbf{r}_1)$ with the boundary condition

\begin{equation}
V_{jl}(\mathbf{r}_b)=\frac{e}{4\pi\varepsilon_0}\int d^3\mathbf{r}_1 \frac{\psi_j^*(\mathbf{r}_1) \psi_l(\mathbf{r}_1)}{\varepsilon |\mathbf{r}_b -\mathbf{r_1}|},
\end{equation}
where $\mathbf{r}_b$ lays within the boundary of the the computational box. The Poisson equation is solved on a grid that covers the dot area. The calculation accuracy is carefully monitored\cite{accuracy} and the configuration-interaction calculation convergence better than $10\mu$ eV is reached for $n=20$.

\subsection{Parameters}
The bulk of our results presented in the following sections are obtained for parameters described below. In the experiment of Ref. \onlinecite{tarucha} an anisotropic InAs self-organized-quantum-dot (SOQD) grown on a $[001]$ GaAs substrate is studied with a confinement potential that is elongated due to the source and drain electrodes placed above the dot. The orientation of the dot with respect to the in-plane crystal directions is not well resolved and in the present work this is taken as an additional parameter which is studied. We take  $R_x=100$ nm as the long and $R_y=60$ nm as the short size of the dot.\cite{tarucha} We take $R_z=10$ nm as a reasonable estimate of the dot height (note that the SOQD has a nominal pyramidal shape\cite{tarucha} with height $20$ nm, but our model is limited to a potential with rectangular shape of vertical cross section). $R_z$ influences the effective strength of the Dresselhaus coupling constant and the orbital effects of the in-plane magnetic field. Results for $R_z>10$ nm are also discussed below.

For the purpose of the present study it is important to notice that the electric field in the growth direction defines the strength of the Rashba coupling. The electric field is influenced by the potential profile within the dot,\cite{fry} the Schottky barrier at the dot/electrode interface, surface charges and applied potentials.\cite{neglfy} The electrostatics of the actual device is complex and its complete description is out of the scope of the present work. Nevertheless, we are able to estimate the external electric field present in the system by considering the stability diagram and the width of the systems.\cite{suptarucha} We estimated the maximal value of the external electric field to be of order $-30$ kV/cm for which the electrons are still present in the dot.\cite{maxfz} From the gate voltage $V_g=-0.4$ V of two-electron spectroscopy we estimated $F_z=-13.6$ kV/cm and this value is used in our numerical calculation. Finally, in this paper we indicate that the ratio of the Rashba coupling strength (that is proportional to $F_z$) to the strength of the Dresselhaus coupling can be extracted from the experimentally measured orientation of the magnetic field for which the SO-induced AC vanishes.

We take the SO coupling parameters as $\alpha = 1.1\;\mathrm{nm}^2$ from Ref. \onlinecite{silva} for the Rashba coupling and $\gamma=26.9\;\mathrm{meVnm}^3$ from Ref. \onlinecite{knap} for the Dresselhaus coupling constant. Material parameters for InAs are adopted from Ref. \onlinecite{willatzen} with values $m^*=0.026$, $g=-17.5$.

\section{Results}

\subsection{Without SO coupling}

\begin{figure}[ht!]
\epsfysize=55mm
                \epsfbox[11 150 580 694] {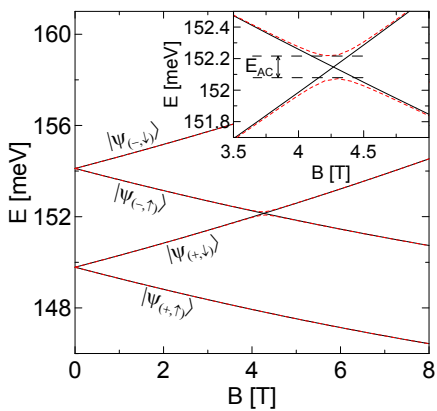}
                 \caption{(Color online) The black solid curves present one electron energy spectrum obtained without spin-orbit coupling for the in-plane magnetic field regardless of the $\phi$ value. The red dashed curves are the energy levels when only Dresselhaus coupling with $\gamma=26.9\;\textrm{meV nm}^3$ is included with magnetic field aligned along the $y$ direction ($\phi=90^\circ$). The inset shows a zoom of the energy levels in the vicinity of the anticrossing.}
 \label{1eD}
\end{figure}

We consider first the dot aligned such that the $x$-axis is oriented along $[100]$ ($y$-axis along $[010]$), namely $\theta=0$. The energy spectrum obtained in the absence of the SO coupling (we take $\alpha=\gamma=0$) for a single-electron anisotropic quantum is presented in Fig. \ref{1eD} by the black solid curves. In the absence of the magnetic field the ground state is doubly degenerate with respect to spin and the spatial wave function is of even symmetry with respect to plane inversions: $\psi(x,y,z)=\psi(-x,y,z)$, $\psi(x,y,z)=\psi(x,-y,z)$, and $\psi(x,y,z)=\psi(x,y,-z)$. We denote the state of even symmetry with respect to all inversions by $|\psi_{+}\rangle$. The first-excited state is a spin-doublet with wave-functions meeting the symmetry conditions: $\psi(x,y,z)=-\psi(-x,y,z)$, $\psi(x,y,z)=\psi(x,-y,z)$, and $\psi(x,y,z)=\psi(x,y,-z)$. We will refer to this state as $|\Psi_{-}\rangle$. The non-zero magnetic field lifts the spin degeneracy splitting of the states of the same parity by the Zeeman energy. The energy levels depicted by the black lines in Fig. \ref{1eD} are obtained regardless of the $\phi$ value in spite of the lateral anisotropy of the dot. Due to the small $R_z$ value and the in-plane alignment of $\mathbf{B}$, no orbital effects of the magnetic field are observed (the influence of the height of the dot is studied in subsection F).

Generally, in the presence of an in-plane magnetic field the Hamiltonian (1), even without SO interaction, does not commute with the plane inversion operators $P_x$ and $P_y$ [defined as $P_xf(x,y,z)=f(-x,y,z)$ and $P_yf(x,y,z)=f(x,-y,z)$]. However, due to the insignificance of the orbital effect of the magnetic field for this flat quantum dot, the parity with respect to reflection through the $x=0$ and $y=0$ plains is approximately preserved (with $\langle P_x \rangle$ and $\langle P_y \rangle$ above $0.97$) even for non-zero $B$. For the following discussion we denote the four lowest-energy states for small magnetic field aligned parallel to the $y$ direction as $|\Psi_{(+,\uparrow)}\rangle$, $|\Psi_{(+,\downarrow)}\rangle$, $|\Psi_{(-,\uparrow)}\rangle$, $|\Psi_{(-,\downarrow)}\rangle$ with corresponding energies $E_{(+,\uparrow)}, E_{(+,\downarrow)}, E_{(-,\uparrow)}, E_{(-,\downarrow)}$ where the arrow denotes the spin state aligned parallel $(\uparrow)$ or antiparallel $(\downarrow)$ to the magnetic field vector.

\subsection{Single type of SO coupling}
Inclusion of the SO interaction lifts the spin polarization of the states and changes the crossing observed between the energy levels of $|\Psi_{(+,\downarrow)}\rangle$, $|\Psi_{(-,\uparrow)}\rangle$ around $B=4.25$ T into an anti-crossing. The inset of Fig. \ref{1eD} shows the anticrossing energy levels for $\phi=90^\circ$ ($\mathbf{B}$ parallel to the $y$ axis) by the red curves when only Dresselhaus coupling with $\gamma=26.9\;\mathrm{meVnm}^3$ is included. We denote the minimal energy difference between the anticrossing levels as $E_{AC}$. For applied parameters we obtain $E_{AC}=146\;\mu$eV. Outside the anticrossing the SO interaction does not modify the energy spectrum in a noticeable way i.e. the black and red curves approximately coincide.
\\
\begin{figure}[ht!]
\epsfysize=39mm
                \epsfbox[11 289 577 552] {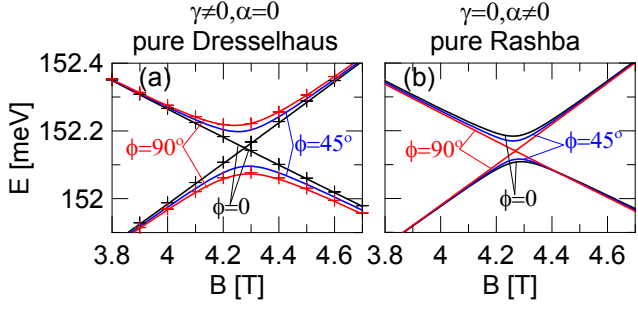}
                 \caption{The energy levels near the anticrossing for pure Dresselhaus (a) and pure Rashba (b) couplings for different $\mathbf{B}$ orientations. The black crosses are the results from a diagonalization of the matrix (\ref{m1}) and the red crosses the results of diagonalization of the (\ref{m2}) matrix. The magnetic field is oriented along the $x$ direction ($\phi=0$) for the black symbols and curves, and along the $y$ direction ($\phi=90^\circ$) for the red symbols and curves.}
 \label{1eRD}
\end{figure}

In the presence of the SO coupling the anticrossing energy levels depend on the orientation of the magnetic field. In Figs. \ref{1eRD}(a) and \ref{1eRD}(b) we plot the energy levels obtained for pure Dresselhaus and pure Rashba interaction, respectively, for three different $\phi$ values. In both cases clear dependence of the anticrossing width $E_{AC}$ is observed with respect to the $\mathbf{B}$ orientation. For pure Dresselhaus coupling the anticrossing is the widest when the magnetic field vector is perpendicular to the $y$ direction ($\phi=90^\circ$) [the red curve in Fig. \ref{1eRD}(a)]. When the field is aligned along the $x$ direction ($\phi=0$) the mixing between levels vanishes [the black curve in Fig. \ref{1eRD}(a)] and there is crossing of levels. With pure Rashba coupling the dependence is opposite -- the anticrossing vanishes when $\mathbf{B}$ is aligned along $y$ and $E_{AC}$ is largest when $\mathbf{B}$ is aligned along $x$.

The direction of the magnetic field for which the mixing between the states disappears can be infered from the analytic form of the SO Hamiltonians utilizing the approximate symmetries of the wave functions of the confined electron. Let us first inspect the case of pure Dresselhaus coupling and remind that for $\theta=0$ the Hamiltonian (\ref{habia3d}) has the same form in the $x,y$ and $z$ coordinate system. Averaging the Hamiltonian (\ref{habia3d}) over the $z$ direction one obtains,
\begin{equation}
\begin{split}
  H_{BIA}^{2D} =& \gamma \langle k^2_z\rangle\left[\sigma_x k_x-\sigma_y k_y\right] + \gamma \left[\sigma_y k_y k^2_x - \sigma_x k_x k^2_y\right]\\
  &+ \gamma \sigma_z \langle k_z\rangle(k^2_y - k^2_x).\label{habia}
\end{split}
\end{equation}

 The second term is the so-called cubic Dresselhaus term which is negligible as long as the height is much smaller than the lateral size of the dot [i.e. until the value of $\langle k^2_x \rangle$ or $\langle k^2_y \rangle$ becomes comparable with $\langle k^2_z \rangle$]. For an infinite quantum well ground-state wave function in the $z$-direction the last term in (\ref{habia}) vanishes\cite{kznegl} and
 \begin{equation}
 \gamma^{2D}=\gamma \langle k_z^2 \rangle=\gamma \left(\pi/R_z\right)^2.\label{gamma2d}
 \end{equation}

The simplified Dresselhaus Hamiltonian takes now the form
\begin{equation}
H_{BIA}^{2D}=\gamma^{2D}\left(\sigma_x k_x-\sigma_y k_y\right)\label{habiasimp}.
\end{equation}

Let us now consider the case of a magnetic field aligned paralel to the $y$ direction. In our basis we include only the low-energy states $|\Psi_{(+,\downarrow)}\rangle$, $|\Psi_{(-,\uparrow)}\rangle$ that exhibit an energy crossing without SO coupling. The matrix of the $H_{BIA}^{2D}$ Hamiltonian limited to this basis is given by,
\begin{widetext}
\begin{equation}
\left(
\begin{array}{cc}
E_{(+,\downarrow)} + \gamma^{2D}\langle\Psi_{(+,\downarrow)}|\sigma_xk_x-\sigma_yk_y|\Psi_{(+,\downarrow)}\rangle & \gamma^{2D}\langle\Psi_{(+,\downarrow)}|\sigma_xk_x-\sigma_yk_y|\Psi_{(-,\uparrow)}\rangle\\
\gamma^{2D}\langle\Psi_{(-,\uparrow)}|\sigma_xk_x-\sigma_yk_y|\Psi_{(+,\downarrow)}\rangle & E_{(-,\uparrow)} + \gamma^{2D}\langle\Psi_{(-,\uparrow)}|\sigma_xk_x-\sigma_yk_y|\Psi_{(-,\uparrow)}\rangle \end{array}\right).
\end{equation}
\end{widetext}
The states $|\Psi_{(+,\downarrow)}\rangle$, $|\Psi_{(-,\uparrow)}\rangle$ are separable into an orbital and a spin part. Due to the action of the Pauli matrices on the states with definite spin one gets,
\begin{widetext}
\begin{equation}
\left(
\begin{array}{cc}
E_{(+,\downarrow)}-\gamma^{2D}\langle\Psi_{(+,\downarrow)}|\sigma_yk_y|\Psi_{(+,\downarrow)}\rangle & \gamma^{2D}\langle\Psi_{(+,\downarrow)}|\sigma_xk_x|\Psi_{(-,\uparrow)}\rangle\\
\gamma^{2D}\langle\Psi_{(-,\uparrow)}|\sigma_xk_x|\Psi_{(+,\downarrow)}\rangle & E_{(-,\uparrow)}-\gamma^{2D}\langle\Psi_{(-,\uparrow)}|\sigma_yk_y|\Psi_{(-,\uparrow)}\rangle \end{array}\right).
\end{equation}
\end{widetext}
For the magnetic field vector aligned parallel to the $y$ direction the components of the momentum operator vector are $k_x=-i \frac{\partial}{\partial x} + \textrm{eB}z, k_y=-i \frac{\partial}{\partial y}, k_z=-i \frac{\partial}{\partial z}$. Due to parity one obtains,

\begin{equation}
\left(
\begin{array}{cc}
E_{(+,\downarrow)} & -i\gamma^{2D}\langle\Psi_{(+,\downarrow)}|\sigma_x\frac{\partial}{\partial x}|\Psi_{(-,\uparrow)}\rangle\\
-i\gamma^{2D}\langle\Psi_{(-,\uparrow)}|\sigma_x \frac{\partial}{\partial x} |\Psi_{(+,\downarrow)}\rangle & E_{(-,\uparrow)} \end{array}\right).\label{m2}
\end{equation}
The non-vanishing off-diagonal matrix elements mix the states $| \Psi_{(+,\downarrow)} \rangle, | \Psi_{(-,\uparrow)} \rangle$ which results in an avoided crossing between the corresponding energy levels. By the red crosses in Fig. \ref{1eRD} we plot numerically calculated eigenvalues of the matrix (\ref{m2}). Note that the crosses and lines are in perfect agreement proving that for our dot with the assumed geometry the $H_{BIA}^{2D}$ is in fact a good approximation to $H_{BIA}$.

Let us now consider the case of a magnetic field aligned parallel to the $x$ ($\phi=0$) direction. In this case the low-energy states which energy levels cross without SO coupling are $|\Psi_{(+,\leftarrow)}\rangle$, $|\Psi_{(-,\rightarrow)}\rangle$, where the arrow denotes the electron spin aligned parallel ($\rightarrow$) and antiparallel ($\leftarrow$) to the magnetic field vector $\mathbf{B}$. The matrix of the $H_{BIA}^{2D}$ Hamiltonian in this two state basis is

\begin{widetext}
\begin{equation}
\left(
\begin{array}{cc}
E_{(+,\leftarrow)}+ \gamma^{2D}\langle \Psi_{(+,\leftarrow)}|\sigma_xk_x-\sigma_yk_y|\Psi_{(+,\leftarrow)}\rangle & \gamma^{2D}\langle\Psi_{(+,\leftarrow)}|\sigma_xk_x-\sigma_yk_y|\Psi_{(-,\rightarrow)}\rangle\\
\gamma^{2D}\langle\Psi_{(-,\rightarrow)}|\sigma_xk_x-\sigma_yk_y|\Psi_{(+,\leftarrow)}\rangle & E_{(-,\rightarrow)}+\gamma^{2D}\langle\Psi_{(-,\rightarrow)}|\sigma_xk_x-\sigma_yk_y|\Psi_{(-,\rightarrow)}\rangle \end{array}\right).
\end{equation}
\end{widetext}
Due to spin one gets,

\begin{widetext}
\begin{equation}
\left(
\begin{array}{cc}
E_{(+,\leftarrow)}+ \gamma^{2D}\langle\Psi_{(+,\leftarrow)}|\sigma_xk_x|\Psi_{(+,\leftarrow)}\rangle & -\gamma^{2D}\langle\Psi_{(+,\leftarrow)}|\sigma_yk_y|\Psi_{(-,\rightarrow)}\rangle\\
-\gamma^{2D}\langle\Psi_{(-,\rightarrow)}|\sigma_yk_y|\Psi_{(+,\leftarrow)}\rangle & E_{(-,\rightarrow)}+\gamma^{2D}\langle\Psi_{(-,\rightarrow)}|\sigma_xk_x|\Psi_{(-,\rightarrow)}\rangle \end{array}\right).\label{m1a}
\end{equation}
\end{widetext}
For the magnetic field aligned along the $x$ direction the components of the momentum operator vector are, $k_x=-i \frac{\partial}{\partial x}, k_y=-i \frac{\partial}{\partial y}, k_z=-i \frac{\partial}{\partial z} + eBy$. All integrals in Eq. (\ref{m1a}) vanish due to the parity of the states and we finally obtain,

\begin{equation}
\left(
\begin{array}{cc}
E_{(+,\leftarrow)} & 0\\
 0& E_{(-,\rightarrow)}\label{m1}
\end{array}\right).
\end{equation}
The matrix (\ref{m1}) consists only of diagonal elements that are equal to the energy of the basis states. Thus the $|\Psi_{(+,\leftarrow)}\rangle$, $|\Psi_{(-,\rightarrow)}\rangle$ states are not mixed by the Dresselhaus coupling in this configuration and there is no anticrossing of energy levels. We plot the eigenvalues of the matrix (\ref{m1}) by the black crosses in Fig. \ref{1eRD}(a).

A similar analysis can be made for the Rashba Hamiltonian (\ref{hasia}). Due to the fact that the analytic form of both Hamiltonians $H_{SIA}$ and $H_{BIA}^{2D}$ are similar, i.e. only the $k_x$ and $k_y$ are swapped (and the coupling constants are different), it is clear that the dependence of AC width on magnetic field direction is opposite -- the mixing between the states vanishes when the magnetic filed is aligned along the $y$ direction.

\subsection{Anisotropy in the presence of both SO couplings}

\begin{figure}[ht!]
\epsfysize=55mm
                \epsfbox[11 183 576 652] {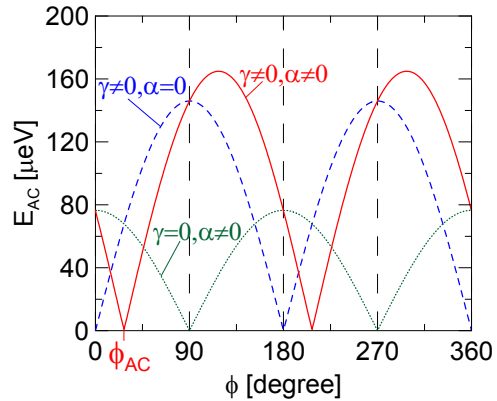}
                 \caption{The spin-orbit-induced anticrossing width $E_{AC}$ for pure Dresselhaus (blue dashed curve), pure Rashba (green dotted curve) and both (red solid line) the interactions present. For $\gamma=0$ the magnetic field is $B=4.268$ T, for the two other cases $B=4.277$ T.}
 \label{1eobr}
\end{figure}

Let us now consider the effect of both Dresselhaus and Rashba coupling. Figure \ref{1eobr} presents the avoided crossing energy $E_{AC}$ as a function of the angle $\phi$ between the $x$ axis and the magnetic field. For pure Dresselhaus (the blue dashed curve in Fig. \ref{1eobr}) and pure Rashba (the green dotted curve in Fig. \ref{1eobr}) coupling the extrema are shifted by $90^\circ$ in agreement with our previous analysis. The curves in Fig. \ref{1eobr} are accurately described by $|\sin(\phi-\phi_{AC})|$ which is the same functional form as the one observed in the experimental work of Ref. \onlinecite{tarucha} in Fig. 3(f) [where the behavior was described by $|\cos(\phi-\phi_0)|$]. Moreover the maximal value of $E_{AC}$ is of the same order as the magnitude observed experimentally. When both SO interactions are present the dependence of the anticrossing width is plotted in Fig. \ref{1eobr} by the red curve. The shape of the latter is the same as for pure Dresselhaus/Rashba coupling with pronounced minima where $E_{AC}$ is zero. When the magnetic field is aligned along the $x$ or $y$ direction the $E_{AC}$ equals the value for pure SO coupling. Note that the maxima are larger than the ones observed for pure couplings and its positions are now shifted and are no longer aligned along the axes of the dot. For $\alpha = 1.1\;\mathrm{nm}^2$ and $\gamma=26.9\;\mathrm{meVnm}^3$ the shift of the dependence is $\phi_{AC}=27.8^\circ$. The latter value can be understood as follows. Let us denote the direction of the magnetic field for which the AC vanishes for pure Dresselhaus and pure Rashba couplings by the vectors $\mathbf{d}_{BIA}$ and $\mathbf{d}_{SIA}$, respectively. Next, we estimate the strength of each interaction. Maximal induced anticrossing width is $E_{AC}^{BIA}=146\;\mu$eV and $E_{AC}^{SIA}=77\;\mu$eV for Dresselhaus and Rashba coupling, respectively. Thus the Dresselhaus interaction is $1.9$ times larger than the Rashba coupling what makes the vector $\mathbf{d}_{BIA}$ $1.9$ times longer than $\mathbf{d}_{SIA}$. Let us denote the magnetic field for which the effect of both spin-orbit couplings is zero by the vector $\mathbf{d}_{BIA+SIA}=\mathbf{d}_{BIA}+\mathbf{d}_{SIA}$. It is easy to show that this vector forms an angle $\phi=27.8^\circ$ with the $x$ axis. Thus when both couplings are present, the effect of the total spin-orbit coupling disappears when the external magnetic field is directed along this vector. In fact that is exactly what we observe in our calculation (see position of the minimum of the dependence depicted with the red curve in Fig. \ref{1eobr}). The formula $|\sin(\phi-\phi_{AC})|$ reflect the fact that the dependency obtained for both SO couplings present can be considered as an absolute value of a sum of the dependencies obtained for pure SO couplings described by $-\cos\phi$ and $\sin\phi$ for pure Rashba and Dresselhaus couplings respectively.

\subsection{Dependence on the quantum dot orientation}

\begin{figure}[ht!]
\epsfysize=43mm
                \epsfbox[19 280 576 553] {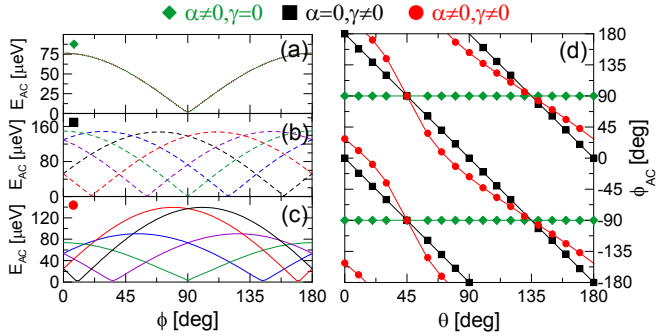}
                 \caption{Avoided crossing energy width as function of the direction ($\phi$) for different orientations of the dot with $\theta=10^\circ$ (black curves), $\theta=30^\circ$ (blue curves), $\theta=45^\circ$ (green curves), $\theta=60^\circ$ (violet curves), $\theta=80^\circ$ (red curves). Results are shown for (a) pure Rashba, pure (b) Dresselhaus and (c) for both the couplings present. (d) The value of the magnetic field angle $\phi_{AC}$ at which $E_{AC}=0$ as a function of the angle $\theta$ for pure Rashba (green diamonds), pure Dresselhaus (black squares) and both couplings present (red dots). The red curves are obtained from Eq. (\ref{vect}).}
 \label{orient}
\end{figure}
Different in-plane orientations of the anisotropic potential of the dot with respect to the crystal host are now considered where the long axis of the dot forms an angle $\theta$ with $[100]$.  In Figs. \ref{orient}(a,b,c) we present the size of the avoided-crossing as a function of the direction of the rotated magnetic field (note that the $\phi$ angle is defined as an angle between the magnetic field vector and the long axis of the dot) for six different orientations of the dot. The dotted curves in Fig. \ref{orient}(a) presents the result obtained for pure Rashba coupling. We observe that the $E_{AC}$ dependencies are exactly the same as in Fig. \ref{1eobr} regardless of the dot alignment. The minimum of the $E_{AC}$ does not change its position and the energy levels are not affected by the orientation of the dot. We show in Fig. \ref{orient}(d) the $\phi_{AC}$ angle for which the $E_{AC}=0$ as a function of the angle $\theta$ by the green diamonds.

For pure Dresselhaus coupling the $E_{AC}$ dependencies [depicted by dashed curves in Fig. \ref{orient}(b)] are shifted as the dot is rotated. For the case studied in previous subsections (where $\theta=0$)  the AC vanished when the magnetic field was aligned along the long axis of the dot ($\phi_{AC}=0$). When the dot is oriented by $\theta=45^\circ$ (long axis oriented along the $[110]$ direction), the anticrossing vanishes when the magnetic field is aligned along the short axis of the dot [see green dashed curve in Fig. \ref{orient}(b)] -- $\phi_{AC}=90^\circ$ -- the same as for pure Rashba case. We plot in Fig. \ref{orient}(d) the angle $\phi_{AC}$ for pure Dresselhaus coupling by the black squares for different orientations of the dot. We find that the angle exhibits a $\phi_{AC}=-2\theta$ dependence [black solid lines in Fig. \ref{orient}(d)]. Moreover we observe that for both cases when only a single type of SO coupling is present the maximal value of the AC width remains unchanged.

\begin{figure}[ht!]
\epsfysize=50mm
                \epsfbox[19 174 576 655] {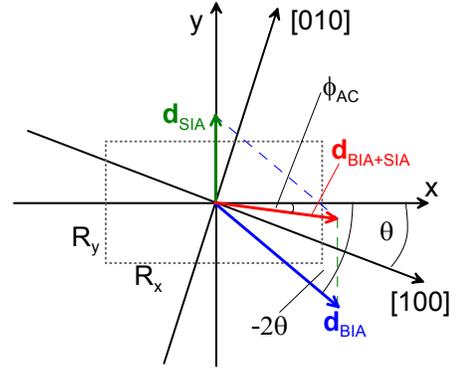}
                 \caption{Schematic of the method of calculation of the angle $\phi_{AC}$ for which the the AC vanishes when both SO couplings are present and the quantum dot (dashed rectangle) is oriented with its long axis forming the angle $\theta$ with the $[100]$ direction. The vectors depict the directions of $\mathbf{B}$ for which $E_{AC}=0$ for pure Rashba coupling (green arrow), pure Dresselhaus coupling (blue arrow) and both couplings present (red arrow). The coordinate system connected with the dot axes and the crystallographic directions is also shown.}
 \label{vectf}
\end{figure}

In Fig. \ref{orient}(c)  we show the results when both SO couplings are present by the solid curves. The maximal values of $E_{AC}$ and the angle $\phi_{AC}$ for which the minima are observed change when the dot orientation is varied. Both facts can be understood similarly as discussed in subsection C. We can justify the $\phi_{AC}$ values considering the orientation of the $\mathbf{d}_{BIA+SIA}=\mathbf{d}_{BIA}+\mathbf{d}_{SIA}$ vector. But now the orientation of the $\mathbf{d}_{BIA}$ vector assigned with Dresselhaus coupling is changed as the dot is rotated, ie. the vector $\mathbf{d}_{BIA}$ forms an angle $-2\theta$ with the long axis of the dot. The rotation of the dot does not change the maximal value of $E_{AC}$ when only a single type of SO coupling is present and the previously derived value for the relative strength of both couplings remains unchanged (and thus also the ratio of the length of the $\mathbf{d}_{BIA}$ and $\mathbf{d}_{SIA}$ vectors). We take $1$ as the length of $\mathbf{d}_{SIA}$ and $1.9$ as the length of $\mathbf{d}_{BIA}$. In Fig. \ref{vectf} we schematically present the considered vectors and the angles they form with the axes of the dot. The angle between the $\mathbf{d}_{BIA+SIA}$ vector (red arrow in Fig. \ref{vectf}) and the $x$ direction can be easily calculated

\begin{equation}
\phi_{AC}=\arctan\left(\frac{1+1.9\sin(-2\theta)}{1.9\cos(2\theta)}\right).\label{vect}
\end{equation}

With the red dots in Fig. \ref{orient}(d) we plot the angle $\phi_{AC}$ obtained from our numerical calculation in the presence of both couplings for different orientations of the dot which agree very well with the values (red curves) obtained from Eq. (\ref{vect}). Along with the changes of the orientation the length of the $\mathbf{d}_{BIA+SIA}$ vector is changed which results in different values of the maximal AC width observed in Fig. \ref{orient}(c).

A systematic study of the value of the $\phi_{AC}$ angle dependence on the SO coupling strengths and the dot alignment is given in subsection G where the two-electron case is studied.

\subsection{Quantum dot with square base}
\begin{figure}[ht!]
\epsfysize=45mm
                \epsfbox[11 261 580 585] {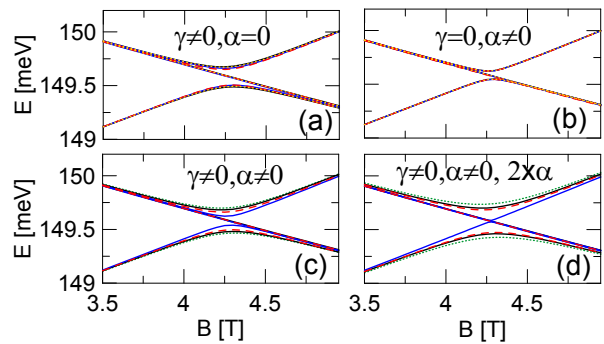}
                 \caption{Energy levels of one-electron quantum dot with square base with $R_x=R_y=100$ nm. The black curves correspond to $\phi=0$, the blue curves to $\phi=45^\circ$, yellow dotted to $\phi=22.5^\circ$ and the red dashed curves to $\phi=90^\circ$. In (c,d) we additionally plot the energy levels obtained for $\phi=135^\circ$ with green dotted curves. (a) Pure Dresselhaus coupling, (b) pure Rashba interaction, (c) both SO interactions are present and (d) both SO interactions are present with $\alpha$ increased by a factor of two.}
 \label{1ekwRD}
\end{figure}

The above discussion was for a lateral anisotropic quantum dot. Now we study the case of a dot with symmetrical base (we assume $R_x=R_y=100$ nm) and $\theta=0$ and investigate if this has an influence on the anisotropy induced by the SO coupling. In the absence of the SO interaction and a magnetic field the first-excited state is spin-doubly degenerate due to parity. The magnetic field lifts the spin degeneracy but the degeneracy due to parity is not removed. The inclusion of a single type of SO interaction induces a repulsion between the energy levels of the ground-state and one of the states from the parity doublet [see the red dashed curves in Fig. \ref{1ekwRD}(a) for the case of pure Dresselhaus coupling and Fig. \ref{1ekwRD}(b) for pure Rashba coupling]. The same configuration of energy levels is obtained regardless of the angle $\phi$. In both Figs. \ref{1ekwRD}(a) and \ref{1ekwRD}(b) the black ($\phi=0$), blue ($\phi=45^\circ$), yellow dotted ($\phi=22.5^\circ$) and red dashed curves ($\phi=90^\circ$) coincide.  The dependence of the energy levels on the magnetic field orientation starts to appear already when the dot is elongated by a factor of $1\%$.

However, when both Rashba and Dresselhaus interactions are present the AC width varies with the rotation of the magnetic field -- Figs. \ref{1ekwRD} (c,d). We observe that the anisotropy is most pronounced when $\alpha$ is increased by a factor of two -- the case when both couplings have comparable strengths.\cite{oblicz1,oblicz2} In such a case when the magnetic field is directed along the diagonal, i.e. $\phi=45^\circ$ [the blue curves in Fig. \ref{1ekwRD}(d)] the anticrossing between the energy levels of the ground-state and both states from the parity-doubled becomes very small.

\subsection{Larger dot height}
\begin{figure}[ht!]
\epsfysize=80mm
                \epsfbox[55 10 550 830] {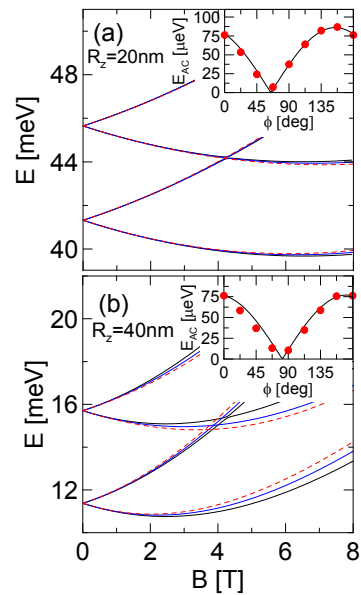}
                 \caption{One-electron energy levels for rectangular-based quantum dot with $R_z=20$ nm (a) and $R_z=40$ nm (b) for $\phi=0$, $\phi=45^\circ$ and $\phi=90^\circ$ plotted with the black solid, blue solid and red dashed curves, respectively. The red dots in the insets to both plots presents the anticrossing width $E_{AC}$ obtained from the energy-spectrum for a given $\phi$ value and the black curves are the fitted $|\sin(\phi-\phi_{AC})|$ dependencies.}
 \label{1ewysoka}
\end{figure}

Let us now return to the case of the quantum dot with rectangular base with $R_x=100$ nm and $R_y=60$ nm. For the previous dot with $R_z=10$ nm no orbital effects from the magnetic field on the energy spectrum was observed [see the black curves in Fig. \ref{1eD}]. However this is no longer true for larger $R_z$ values. This can be seen from Figs. \ref{1ewysoka}(a) and \ref{1ewysoka}(b) where we plot the energy levels of a quantum dot with height $R_z=20$ nm and $R_z=40$ nm, respectively, in the presence of SO coupling (with both SO interactions present). The energy levels depend on the magnetic field orientation even outside the anticrossing region. This is due to the elongation of the confinement potential in the $x$ direction. The SO-induced anticrossing is shifted to lower magnetic fields as the value of the angle $\phi$ becomes closer to $\phi=90^\circ$ [this is analogous to the experimental observation -- compare with Fig. S7(a) from Ref. \onlinecite{suptarucha}]. We calculated the anticrossing widths for different values of $\phi$ and plot them as red dots in the insets of Fig. \ref{1ewysoka}(a,b). Then we fitted the points with the function $A|\sin(\phi-\phi_{AC})|$ where $A=86\;\mu$eV, $\phi_{AC}=65^\circ$ for $R_z=20$ nm and $A=77\;\mu$eV, $\phi_{AC}=82^\circ$ for $R_z=40$ nm. Notice the agreement between the fitted curve and the data points. From this fact we conclude that in spite of the presence of orbital effects, previously found dependence of the anticrossing width on the angle $\phi$ still holds, but with modified $A$ and $\phi_{AC}$ values. The latter fact can be attributed to the reduction of the Dresselhaus coupling strength. This can be accounted for by considering the Dresselhaus coupling Hamiltonian (\ref{habiasimp}) in which the coupling strength decreases as $(1/R_z)^2$. In the calculation performed for pure Dresselhaus interaction we obtain the maximal $E_{AC}$ values $146\;\mu$eV, $37\;\mu$eV and $11\;\mu$eV for $R_z=10$ nm, $R_z=20$ nm and $R_z=40$ nm, respectively. For increased dot height the obtained $E_{AC}$ values decrease approximately as $(1/R_z)^2$ with the largest discrepancy for large $R_z$ value (i.e. when the approximation of the coupling strength by Eq. (\ref{gamma2d}) becomes inaccurate). The decrease of the Dresselhaus coupling strength for increased height of the dot results in a shift of the $E_{AC}$ dependency on $\phi$ towards the one obtained for a flat quantum-dot with only Rashba interaction present (compare the black curve in the inset of Fig. \ref{1ewysoka}(b) with the green dotted curve in Fig. \ref{1eobr}) -- $\phi_{AC}$ becomes close to $90^\circ$. Also the maximal $E_{AC}$ value becomes closer to the one obtained for pure Rashba coupling -- $A$ tends to $77\;\mu$eV with increasing $R_z$. The shift in the $\phi_{AC}$ value [see insets of Figs. \ref{1ewysoka} (a,b)] can be understood from the relative strengths of the Rashba and Dresselhaus coupling as discussed in section III. C.

\subsection{Two electron results}
In a recent experiment [\onlinecite{tarucha}] the ground-state and excited states were measured provided that the latter entered into a finite but narrow transport window determined by the voltages applied to the source and drain electrodes. The avoided crossings that appear for a single-electron in the excited part of the spectrum, which we described above, were outside the transport window.

In the two-electron regime and in the absence of both the magnetic field and the SO interaction the ground-state is a spin-singlet and the first excited state is a spin-triplet. Under the presence of an external magnetic field the ground-state singlet energy crosses the triplet energy. When we turn on the SO coupling it induces an avoided crossing between the states of opposite spin which was well resolved in the experiment [\onlinecite{tarucha}].

\begin{figure}[ht!]
\epsfysize=60mm
                \epsfbox[11 147 590 705] {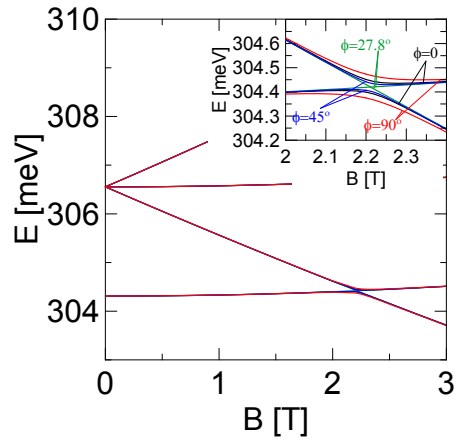}
                 \caption{Two-electron energy spectrum in the presence of both Rashba and Dresselhaus SO coupling, for angles $\phi=0$ (black curves), $\phi=27.8^\circ$ (green curves), $\phi=45^\circ$ (blue curves) and $\phi=90^\circ$ (red curves). The inset shows the energy levels in the vicinity of the anticrossing. The results are obtained for $\theta=0$.}
 \label{2eRD}
\end{figure}

 Similarly to the one-electron case the SO coupling is responsible for changes in the size of the anticrossing energy when the orientation of the magnetic field is varied. Figure \ref{2eRD} presents the low-energy spectrum of the two-electron quantum dot in the presence of both Rashba and Dresselhaus coupling for a dot aligned with its long axis along the $[100]$ direction ($\theta=0$). In the inset we plot the energy levels in the vicinity of the avoided crossing. The anticrossing vanishes for exactly the same angle $\phi_{AC}=27.8^\circ$ as for the one-electron case discussed above (see the green curves in the inset of Fig. \ref{2eRD}).

\begin{figure}[ht!]
\epsfysize=100mm
                \epsfbox[28 14 552 820] {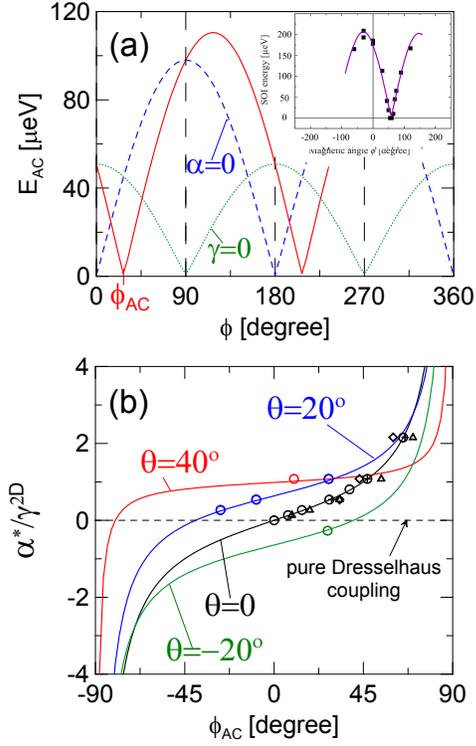}
                 \caption{(a) Width of the singlet-triplet avoided crossing as a function of the angle $\phi$ for pure Dresselhaus coupling (blue dashed curve), pure Rashba coupling (green dotted curve) and for both couplings present (red solid curve). The inset shows the experiment results (symbols) of Ref. \onlinecite{tarucha} together with the result of the present calculation (purple curve) with $\gamma=29.58\;\mathrm{meVnm}^3$ and $\alpha=4.731\;\mathrm{nm}^2$. The magnetic field is $B=2.211$ T for pure Rashba coupling and $B=2.209$ T for all the other cases. The dot is aligned with its long axis along $[100]$ i.e. ($\theta=0$). (b) The $\phi_{AC}$ value for different strength ratios of the Rashba and Dresselhaus coupling for four different orientations of the dot. The symbols present the results of our numerical calculation and the curves represent the analytical result given by Eq. (\ref{fin}). The circles show the results obtained for $R_y=60$ nm, $R_z=10$ nm, the crosses for $R_y=30$ nm, $R_z=10$ nm, the triangles for $R_y=20$ nm, $R_z=10$ nm, and the diamonds for $R_y=60$ nm, $R_z=20$ nm. In all cases $R_x$ is $100$ nm.}
 \label{2efin}
\end{figure}

In Fig. \ref{2efin}(a) we plot the angular dependence of the anticrossing width $E_{AC}$ for pure Dresselhaus, pure Rashba and when both couplings are present by the blue dashed, green dotted and solid red curves, respectively. Notice that all three dependencies have the same shape as for the case of the one-electron considered in subsection B (compare with Fig. \ref{1eobr}), only the maximal $E_{AC}$ values are about $1.5$ times smaller.

As was presented in subsection C for the anisotropic quantum dot the angle $\phi_{AC}$ depends on the relative strength of both SO interactions and the in-plane orientation of the dot (explained in subsection D). On the other hand the $\phi_{AC}$ value can be measured experimentally\cite{tarucha} and the orientation of the quantum dot with respect to the crystal directions can be obtained by inspecting the facets of the dot. This opens the possibility to employ such a measurement to evaluate the relative strength of the Rashba and Dresselhaus couplings for a dot with given orientation with respect to the crystal host. Let us define the strength ratio of the SO interactions as the ratio of the effective coupling constants $\alpha^*$ and $\gamma^{2D}$. The Rashba coupling strength denoted with $\alpha^*$ is calculated as $\alpha^*=-\alpha\left[ \frac{\partial V}{\partial z}\right]=-\alpha|e|F_z$ and the Dresselhaus coupling $\gamma^{2D}$ is obtained from Eq. (\ref{gamma2d}).

We previously derived the angle $\phi_{AC}$ for given relative strength of the SO couplings for a given orientation of the dot [see Eq. (\ref{vect})]. Let us substitute the $1/1.9$ value by $\alpha^*/\gamma^{2D}$ in Eq. (\ref{vect})  from which we obtain
\begin{equation}
\begin{split}
\frac{\alpha^*}{\gamma^{2D}}=-\frac{\alpha}{\gamma}\frac{|e|F_z R_z^2}{\pi^2}=\cos(2\theta)\left[\tan(\phi_{AC})-\tan(-2\theta) \right]\label{fin}.
\end{split}
\end{equation}
This function is shown in Fig. \ref{2efin}(b) by the solid lines for different orientations of the quantum dot. With the black symbols we mark the angle $\phi_{AC}$ obtained from our numerical calculations for dots with different geometries (see figure caption) with $\theta=0$ for different SO coupling strengths. For such case with $\theta=0$ (the dot oriented with its long axis along $[100]$) and pure Dresselhaus coupling ($\alpha^*/\gamma^{2D}=0$) we obtain $\phi_{AC}=0$. When the Rashba coupling strength is increased the points move toward the angle $\phi_{AC}=90^\circ$ obtained for pure Rashba SO coupling. The green, red and blue symbols in Fig. \ref{2efin}(b) are the $\phi_{AC}$ values obtained from our two-electron numerical calculation for different orientation of the quantum dot.

In the above discussion we assume that $\alpha^*=-\alpha |e| F_z$ and $\gamma^{2D}=\gamma \pi^2/R_z^2$ describe the strength of the spin-orbit interactions. For the Rashba coupling given by the Hamiltonian (\ref{hasiafz}) [i.e. when an electric field is only present in the growth direction] the above $\alpha^*$ expression is valid regardless of the dot geometry. However, due to the fact that $\gamma^{2D}$ originates from the Hamiltonian (\ref{habia}) it describes the strength of the Dresselhaus coupling correctly only when the cubic term $\gamma \left[\sigma_y k_y k^2_x - \sigma_x k_x k^2_y\right]$ is negligible, which is the case when $R_x,R_y\gg R_z$ and when the term with $\langle k_z \rangle$ is close to zero i.e. for a dot with limited height.\cite{kznegl} All the symbols in Fig. \ref{2efin}(b) approximately coincide with the dependency given by Eq. (\ref{fin}). Discrepancy is seen in the limit of a narrow dot with $R_y=20$ nm (the triangles) and for increased height of the dot for $R_z=20$ nm (diamond symbols). We conclude that for anisotropic quantum dots with limited height the $\alpha^*/\gamma^{2D}$ ratio is a good measure of the relative strength of the Rashba and the Dresselhaus spin-orbit couplings which can be estimated from analytic expression (\ref{fin}).

The experiment of Ref. \onlinecite{tarucha} found $\phi_{AC}=59^\circ$ and we can use Eq. (\ref{fin}) to calculate the relative strength of the SO interactions. However, as the orientation of the anisotropic potential of the dot with respect to the crystal directions was not resolved in the experiment we need to assume a value for $\theta$. We take $\theta=0$ and by matching the absolute value of the SO coupling constants (through the maximal value of $E_{AC}$) we obtained $\alpha^*/\gamma^{2D}\simeq1.66$ by fitting the experimentally measured values for the AC width with our simulation results. In the inset to Fig. \ref{2efin}(a) we plot our results (purple curve) for the SO coupling constants $\gamma=29.58\;\mathrm{meVnm}^3$, $\alpha=4.731\;\mathrm{nm}^2$ together with the data points from Ref. \onlinecite{tarucha}. However as the relation between the crystal directions and the long axis of the dot is not known the fit only proves the validity of the discussed process behind the anisotropy and not the exact value of the ratio $\alpha^*/\gamma^{2D}$. Moreover as the electrostatics of the actual device is complex the presented result is not the \emph{exact} simulation of the experiment. Therefore, we present in Table I the strength ratios for different orientations of the dot. Note that Eq. (\ref{fin}) does not allow to calculate the relative strength of the couplings for a dot aligned with long axis exactly along $[110]$ or $[1\overline{1}0]$. In a such configuration for pure Dresselhaus as well as for pure Rashba coupling the AC vanishes for $\phi_{AC}=90^\circ$ [compare dotted curves in Fig. \ref{orient}(a) with green-dotted curve in Fig. \ref{orient}(b)] and by that for both couplings present simultaneously the minimum of the $E_{AC}$ dependence on $\phi$ is not shifted irrespective of the coupling strength ratio.

\begin{table}
\begin{tabular}{|c|c|}
  \noalign{\hrule height 0.7pt}
  $\theta$ & $\alpha^*/\gamma^{2D}$ \\
  \noalign{\hrule height 0.7pt}
  0 & 1.66 \\\hline
  $40^\circ$ & 1.27 \\\hline
  $45^\circ$ & --\\\hline
  $75^\circ$ & -0.94\\\hline
  $90^\circ$ & -1.66\\
  \hline

\end{tabular}
 \label{tt1}
  \caption{Calculated strength ratios of the SO couplings for $\phi_{AC}=59^\circ$ and different orientations of the dot.}
\end{table}

\section{Discussion}
In the present paper we discussed the avoided-crossings of energy levels as induced by the presence of different SO couplings. Only for the case of a square-based quantum-dot [see Figs. \ref{1ekwRD}(a,b)] the dependence of AC width as function of the magnetic field direction was observed solely for both couplings present with comparable strength. This result is related to those of Ref. \onlinecite{oblicz1} where the spin-splitting of {\it single}-electron energy levels in strictly {\it two}-dimensional {\it circular} quantum dots in the presence of a {\it small} in-plane magnetic field (before the crossings/avoided-crossings appear) was calculated. When Dresselhaus and Rashba coupling strengths are equal a well known high symmetric case is found which is beneficial for many spintronics applications.\cite{psh,nonsfet} For that special case the energy spectrum is not affected by SO interaction effects and the spin in the [110] direction is strictly defined. The Zeeman interaction lifts this symmetry and results in a spectrum that depends on the orientation of the magnetic field as discussed in Ref. \onlinecite{oblicz1}. Since for equal coupling strengths the spins in the $[110]$ direction are well defined, the Zeeman interaction for $\mathbf{B}$ oriented along $[110]$ does not produce any AC between energy levels of spin-orthogonal states [see the blue curve in Fig. \ref{1ekwRD}(d)].

On the other hand, in the presence of a vertically oriented magnetic field, the size of the Zeeman interaction induced lifting of the symmetry depends on the in-plane orientation \cite{oblicz2} and also the width\cite{prab} of the dot what results in changes in both the AC width and the effective $g$-factor which are solely observed when both SO interactions are present with comparable strength. However, changing the dot orientation is hardly achievable experimentally and therefore in the present work we considered an anisotropy that can be probed by changing the orientation of the magnetic field.

In the present work we investigated the anisotropic dependence of the avoided-crossing width that occurs even for a single type of SO coupling [see Figs. \ref{1eRD}(a) and \ref{1eRD}(b)]. This effect is strictly connected both with the elongation of the confinement potential and the in-plane alignment of the magnetic field [see the discussion in subsection B]. The exact shape of the confinement potential is not important for the studied phenomena which is a generic propriety of a spin-orbit-coupled quantum dot. In our analysis we indicated the trends that determined the dependence of $E_{AC}$ on $\phi$, in particular the dependence on the dot geometry [for the dot with increased height and for different lateral sizes of the dot -- see the black symbols in Fig. \ref{2efin} that in spite of the different geometries of the dot still undergo the same analytical dependence Eq. (\ref{fin})] or the orientation of the quantum dot with respect to the crystallographic directions (which influences the position of the minima of $E_{AC}$ purely due to Dresselhaus coupling -- see discussion in section III. D).

The present study shows that for an elongated quantum dot with pure Rashba coupling the anticrossing vanishes always when the magnetic field is aligned along the short axis of the dot [see the minima of the dotted curves in Figs. \ref{1eobr}, \ref{orient}(a) and \ref{2efin}(a)]. Only the presence of Dresselhaus coupling can result in a $\phi_{AC}$ value that is different from $90^\circ$. The magnetic field direction ($\phi_{AC}=59^\circ$) for which the anticrossing vanished in the experiment of Ref. \onlinecite{tarucha} suggests both SO couplings are present, contrary to the argumentation provided in Ref. \onlinecite{tarucha}. The authors suggested that the Dresselhaus coupling would not induce mixing between the two lowest-energy states due to their well defined and different values of the total angular momentum $J_-=L-S$ in a high magnetic field. However we found, that due to the in-plane alignment of the magnetic field\cite{angu} the Dresselhaus coupling in fact induces avoided-crossings in the energy spectrum of a flat quantum dot [see Fig. \ref{1eD}] and leads also to a shift in the dependence of the AC width on the magnetic field direction [see Fig. \ref{1eRD}, Fig. \ref{orient}(a,b) and insets to Fig. \ref{1ewysoka}].

\section{Summary and conclusions}
We presented a study of the energy spectrum of one and two-electron spin-orbit-coupled three-dimensional quantum dots in the presence of an external in-plane magnetic field. We found that the size of the avoided-crossings in one- and two-electron energy-spectrum oscillates as a function of the orientation of the magnetic field. The oscillatory behavior could accurately be described  by $|\sin(\phi-\phi_{AC})|$ which agrees with recent excited-state spectroscopy measurements performed on InAs gated self-organized-quantum dot.\cite{tarucha}

For a quantum dot which is elongated in the $[100]$ direction and when only a single type of SO coupling is present the avoided crossing vanishes for $\phi_{AC}=0$ ($\phi_{AC}=90^\circ$), i.e. when the magnetic field is aligned parallel to the long (short) axis of the dot for Dresselhaus (Rashba) coupling. We explain this behavior as a consequence of parity- and spin-dependent mixing of the states caused by the SO interaction. When both couplings are present the $\phi_{AC}$ value varies between $0$ and $90^\circ$ and the ratio of the relative strength of the interactions follows a $\tan(\phi_{AC})$ dependence. The change of the in-plane dot orientation results in a change of $\phi_{AC}$ which is observed only when Dresselhaus coupling is present. We show that the experimentally measured $\phi_{AC}$ value\cite{tarucha} along with the knowledge of the orientation of the dot can be used to determine the ratio of the strengths of the individual SO interactions in case of anisotropic quantum dots.

\section*{Acknowledgements}
The authors thank S. Takahashi for helpful discussions. This work was supported by the "Krakow Interdisciplinary PhD-Project in Nanoscience and Advanced Nanostructures" operated within the Foundation for Polish Science MPD Programme co-financed by the EU European Regional Development Fund, the Project No. N N202103938 supported by Ministry of Science an Higher Education
(MNiSW) for 2010–2013 and the Belgian Science Policy (IAP). W.P.  has been partly supported by the EU Human Capital Operation Program, Polish Project No. POKL.04.0101-00-434/08-00. Calculations were performed in ACK\---CY\-F\-RO\-NET\---AGH on the RackServer Zeus.


\begin{thebibliography}{00}
\bibitem{neglfy} The in-plane electric field $F_y$ originating from the potential difference between the source-drain electrodes in Ref. \onlinecite{tarucha} is of order of hundreds of V/cm which is $50$ times smaller than the electric field in the growth direction. The effect of $F_y$ on the energy levels was found to be neglible.
\bibitem{accuracy} The accuracy of the two-electron calculation for an infinite quantum well confinement potential has been checked by comparing the present calculation in the two dimensional limit (in the $z$ direction we use an infinite quantum well with width $1$ nm) with the two-dimensional calculation on the grid showing an accuracy better than $1\;\mu\textrm{eV}$ with ground state energy $E=3.574$ meV. For parabolic confinement we compared the present computation scheme with center-of-mass approach obtaining differences in the energy of order tenth of $\mu \textrm{eV} $ with the ground-state energy $E=248.15$ meV.
\bibitem{maxfz} Evaluated from gate voltage $V_g=-0.8$ V for which the first current peak in tunneling spectroscopy is observed [see Ref. \onlinecite{suptarucha}].
\bibitem{kznegl} The energetic contribution to the AC calculated for $R_z=10$ nm from the second and third term in Eq. (\ref{habia}) is $4\;\mu$ev and $9\;\mu$eV respectively and does not change when the dot height is increased to $R_z=40$ nm.
\bibitem{angu} In a sufficiently strong magnetic field the diamagnetic term of the Hamiltonian becomes stronger than the confinement potential, and in consequence the Hamiltonian eigenstates become identical to Fock-Darwin states. In particular, the ground-state and the first excited state acquire angular momenta $0$ and $-1$, respectively. However, in the problem studied in this paper, including strong confinement in the growth direction and in plane magnetic field orientation, this limit is not reached within the considered range of B. Namely, in the absence of SO coupling at $B=10$ T we obtain mean values of the anular momentum $0.1, 0.1, 0.1, 0.09$ and $0.8, 0.7, 0.6, 0.4$ for the ground and first excited states for $R_z=10$ nm and $R_z=40$ nm respectively.
\end{thebibliography}
\end{document}